\def\mycmd{2} 

\if\mycmd1
\documentclass[draftclsnofoot, english, a4paper, onecolumn, 10pt]{IEEEtran}
\else
\documentclass[journal]{IEEEtran}
\fi

\usepackage[T1]{fontenc}
\usepackage{geometry}
\usepackage{booktabs}
\usepackage{makecell}
\if\mycmd1
\else
\fi

\usepackage{color}
\usepackage{babel}
\usepackage{verbatim}
\usepackage{float}
\usepackage{amsmath}
\usepackage{amssymb}
\usepackage{graphicx}
\usepackage[table]{xcolor}
\usepackage[utf8]{inputenc}
\usepackage{lipsum}
\usepackage{mathtools}
\usepackage{cuted}
\usepackage{multirow}
\usepackage{array}
\usepackage{latexsym}
\usepackage{stfloats}
\usepackage{algorithm}
\usepackage{cite}
\usepackage[labelformat=parens,subrefformat=parens,caption=false]{subfig}
\usepackage[noend]{algpseudocode}

\usepackage{tikz}
\usetikzlibrary{positioning}

\makeatletter

\floatstyle{ruled}
\newfloat{algorithm}{tbp}{loa}
\providecommand{\algorithmname}{Algorithm}
\floatname{algorithm}{\protect\algorithmname}

\makeatother

\begin{document}

\title{Enhancing Sum-Rate Performance in Constrained Multicell Networks: A Low-Information Exchange Approach}

\author{Youjin Kim, \IEEEmembership{Member, IEEE}, Jonggyu Jang, \IEEEmembership{Member, IEEE}, and Hyun Jong Yang, \IEEEmembership{Member, IEEE} \thanks{Y. Kim is with Communication Processor R\&D Team, Mobile eXperience, Samsung Electronics, Suwon, Korea (e-mail: youjin8022@gmail.com). J. Jang and H. J. Yang (corresponding author) are with the Dept. of Electrical Engineering, Pohang University of Science and Technology (POSTECH), 37673, Republic of Korea (e-mail: \{jgjang, hyunyang\}@postech.ac.kr).}}

\maketitle
\begin{abstract}
Despite the extensive research on massive MIMO systems for 5G telecommunications and beyond, the reality is that many deployed base stations are equipped with a limited number of antennas rather than supporting massive MIMO configurations. Furthermore, while the cell-less network concept, which eliminates cell boundaries, is under investigation, practical deployments often grapple with significantly limited backhaul connection capacities between base stations. This letter  explores techniques to maximize the sum-rate performance within the constraints of these more realistically equipped multicell networks. We propose an innovative approach that dramatically reduces the need for information exchange between base stations to a mere few bits, in stark contrast to conventional methods that require the exchange of hundreds of bits. Our proposed method not only addresses the limitations imposed by current network infrastructure but also showcases significantly improved performance under these constrained conditions. 
\end{abstract}

\IEEEpeerreviewmaketitle{}

\section{Introduction}

\label{sec:introduction}

Dense small cell networks are considered one of the key technologies for dealing with excessive data traffic.
In dense small cell network situations, interference management between cells, and proper power allocation, are essential for increasing the sum-rate.
Massive Multi-input multi-output (MIMO) employed at the transmitter can be the solution to improve spectral efficiency gain \cite{L_Lu14_JSTSP,H_Tabassum16_TCom}.
However, in the pervasive mobile networks \cite{TR38.802}, the number of transmit antennas even at base stations (BSs) is often limited by up to 8.
Although a notable sum-rate gain over the conventional distributed beamforming can be achieved with global channel state information (CSI) \cite{S_He14_Elsevier,E_Larsson08_JSAC}, the capacity of a wireless direct link is limited by 10 to 100 Mbps in conventional mobile networks.

Locally optimal solutions are obtained through an iterative approach \cite{Q_Shi11_TSP}, where beamforming vectors, receive equalizers, and weight coefficients are designed iteratively between the transmitters and receivers.
However, the scheme proposed in \cite{Q_Shi11_TSP} requires excessive vector information exchange and the amount of information exchange is proportional to the the number of required iteration number.
In \cite{Y_Kim20_TVT}, non-iterative approach with limited scalar information exchange between BSs is proposed for maximizing the sum-rate without considering channel variation with time.
The information exchange in \cite{Y_Kim20_TVT} is required for every time slot when a time-varying channel is considered, and hence, excessive amount of scalar information is expected as the considered time slots become longer. 

Recently, applying deep neural networks (DNNs) has shown great potentials for improving the sum-rate of wireless communication systems \cite{H_Huang18_Access, W_Xia20_TCom, H_Sun18_TSP, C_Sun20_Access, H_Kwon21_Access}.
In \cite{H_Huang18_Access} and \cite{W_Xia20_TCom}, the beamforming vectors for multiuser MIMO networks with single BS are derived based on unsupervised learning and a hybrid method of supervised and unsupervised learning, respectively, to maximize the sum-rate.
In \cite{H_Sun18_TSP}, a DNN is used to solve power control problem maximizing the sum-rate for the single-input single-output network.
However, in \cite{H_Huang18_Access, W_Xia20_TCom, H_Sun18_TSP}, multicell multiantenna networks are not considered. 
In \cite{C_Sun20_Access, H_Kwon21_Access}, multicell multiantenna networks are considered for designing the beamforming vectors to maximize the sum-rate by selecting the proper combining factors between the traditional maximum-ratio transmission and zero-forcing with a DNN.
Though a DNN based on supervised learning is considered for \cite{H_Kwon21_Access}, the proposed scheme in \cite{H_Kwon21_Access} requires global CSI and more excessive computation is required to get the optimal solution of the sum-rate maximization problem, which is the training data for the supervised learning, as the numbers of cells and antennas are increased.
In \cite{C_Sun20_Access}, a DNN based on a deep Q-network (DQN) is considered, but the maximum sum-rate of the proposed scheme in \cite{C_Sun20_Access} is lower than that of the scheme which requires only local CSI \cite{R_Zakhour09_ITG}.

In this correspondence, a beamforming vector design based on a DQN is proposed for time-varying multicell multi-input single-output (MISO) downlink channels with a few scalar information exchange and local CSI.
The salient contributions of our work are threefold:
\begin{itemize}
    \item \textbf{Weight coefficients design utilizing the DQN:} In the proposed scheme, a DQN is utilized to select the proper weight coefficients of the combination of weighted signal-to-leakage-plus-noise ratio (WSLNR) and weighted generating-interference (WGI) \cite{Y_Kim20_TVT} which maximizes the sum-rate in a distributed way. Thus, the optimization of the beamforming vectors to maximize the sum-rate, which is known as an NP-hard problem, is simplified as the weight coefficients design with the discretized action space.
    \item \textbf{Adaptation in time-varying channels:} Regarding the mobility of the users and the consequential Doppler effect, time-varying channels are considered for the proposed scheme. In the proposed scheme, only one BS updates its weight coefficients in each time slot. Thus, only the information exchange about the updated weighted coefficients is required in each time slot.
    \item \textbf{Small amount of information exchange:} Due to the effective algorithm with the discretized action space, the proposed scheme can achieve notable sum-rate performance compared to the existing schemes with only a few bits of required information exchange. Simulation results show that the effectiveness and feasibility of the proposed algorithm with the DQN.
\end{itemize}

\section{System model and Proposed Beamforming Protocol}

\subsection{System Model}
\label{sec:System}
We consider downlink MISO interference channels. It is assumed that there are $N_C$ cells and each cell is composed of a BS with $N_T$ transmit antennas and a user with a single antenna. In this correspondence, a time-varying channel is considered and the channel vector from the $i$-th BS (referred to as BS $i$ henceforth) to the user in the $j$-th cell (referred to as user $j$ henceforth) over the time slot $t$ is denoted by $\mathbf{h}_{ij,t}\in\mathbb{C}^{N_{T}\times1}$. Since channel reciprocity is assumed, it is assumed that all the BSs have the information about the outgoing channels, i.e., BS $i$ has the information of $\mathbf{h}_{ij,t}$, $j\in\left\{0,\ldots,N_{C}-1\right\} \triangleq\mathcal{N}_{C}$.

With multiple antennas at the transmitters, BS $i$ designs its beamforming vector $\mathbf{w}_{i,t}\in\mathbb{C}^{N_{T}\times1}$ over the time slot $t$.
Considering the power limitation of BSs, it is assumed that $\left\Vert \mathbf{w}_{i,t}\right\Vert ^{2}\leq1$, $i\in\mathcal{N}_C$. Then, the received signal at user $i$ over the time slot $t$ is denoted by
\begin{equation}\label{eq:receivedsignal}
y_{i,t}=\underbrace{\mathbf{h}_{ii,t}^{H}\mathbf{w}_{i,t}x_{i}}_{\textrm{desired\,signal}}+\underbrace{\sum_{k\in\mathcal{N}_C \setminus\{i\}}\mathbf{h}_{ki,t}^{H}\mathbf{w}_{k,t}x_{k}}_{\textrm{intercell\,interference}}+z_{i,t},
\end{equation}
where $x_{l}$ is the unit-variance transmit symbol at BS $l$, $l\in\mathcal{N}_{C}$, and $z_{i,t}$ is the additive white Gaussian noise (AWGN) at user $i$ over the time slot $t$ with zero-mean and variance of $N_{0}$. 
With the received signal in \eqref{eq:receivedsignal}, the corresponding SINR is denoted by
\begin{equation}
\gamma_{i,t}=\frac{\left|\mathbf{h}_{ii,t}^{H}\mathbf{w}_{i,t}\right|^{2}}{{\displaystyle \sum_{k\in\mathcal{N}_C\setminus\{i\}}}\left|\mathbf{h}_{ki,t}^{H}\mathbf{w}_{k,t}\right|^{2}+N_{0}},\label{eq:SINR_definition}
\end{equation}
and the achievable sum-rate is expressed by
\begin{equation} \label{eq:sum_rate}
R_t=\sum_{i=1}^{N_{C}}\log(1+\gamma_{i,t}).
\end{equation}

\subsection{Proposed Beamforming Formulation}\label{sec:bf_formulation}

It is well known the maximization of sum-rate \eqref{eq:sum_rate} is NP-hard and requires global CSI.
According to \cite{Y_Kim20_TVT,E_Bjornson10_TSP}, the solution to the sum-rate maximization problem can be obtained by solving the max-WSLNR problem with the appropriate weight coefficients.

Let us denote the weight coefficient for the channel gain from BS $i$ to user $j$ over the time slot $t$ by $\beta_{ij,t}\geq0$, and the vector of $\beta_{ij,t}$, $j\in\mathcal{N}_C$, by $\boldsymbol{\beta}_{i,t}$.
Then, according to \cite{Y_Kim20_TVT}, the beamforming vector in the max-WSLNR problem can be represented as the function of the weight coefficients as
\if 1\mycmd
\begin{equation}\label{eq:WSLNR_beamforming_vector_design}
    \mathbf{w}_{1,t}(\boldsymbol{\beta}_{i,t}) = \arg_{\mathbf{w}_{i,t}, \left\Vert \mathbf{w}_{i,t}\right\Vert ^{2}\leq1}\max \frac{\beta_{ii,t}\left|\mathbf{h}_{ii,t}^{H}\mathbf{w}_{i,t}\right|^2}{\sum_{j\in\mathcal{N}_{C}\setminus\{i\}}\beta_{ij,t}\left|\mathbf{h}_{ij,t}^{H}\mathbf{w}_{i,t}\right|^2+N_0},
\end{equation}
\else
\begin{align}\label{eq:WSLNR_beamforming_vector_design}
    & \mathbf{w}_{1,t}(\boldsymbol{\beta}_{i,t}) \\ & = \arg_{\mathbf{w}_{i,t}, \left\Vert \mathbf{w}_{i,t}\right\Vert ^{2}\leq1}\max \frac{\beta_{ii,t}\left|\mathbf{h}_{ii,t}^{H}\mathbf{w}_{i,t}\right|^2}{\sum_{j\in\mathcal{N}_{C}\setminus\{i\}}\beta_{ij,t}\left|\mathbf{h}_{ij,t}^{H}\mathbf{w}_{i,t}\right|^2+N_0}, \nonumber
\end{align}
\fi
and the sum-rate \eqref{eq:sum_rate} also can be represented as the function of the weight coefficients.
Here, the weights should be jointly optimized to maximize the sum-rate as follow:
\begin{align} \label{eq:optimal_beta}
    \left( \boldsymbol{\beta}_{1,t}^*, \ldots, \boldsymbol{\beta}_{N_C,t}^*\right) = \arg_{\boldsymbol{\beta}_{1,t}, \ldots, \boldsymbol{\beta}_{N_C,t}} \max R_t \left( \boldsymbol{\beta}_{1,t}, \ldots, \boldsymbol{\beta}_{N_C,t}\right)
\end{align}

The problems \eqref{eq:WSLNR_beamforming_vector_design} and \eqref{eq:optimal_beta} are coupled with each other and all the vectors of the weight coefficients, $\boldsymbol{\beta}_{i,t}$, $i\in\mathcal{N}_C$, need to be jointly optimized, and thus global CSI is required to solve these problems. However, due to the limited direct link capacity, we aim to maximize the sum-rate by designing the proper weight coefficients in \eqref{eq:WSLNR_beamforming_vector_design} with local CSI and some exchanged scalar information between BSs in a distributed way utilizing DQN.

To begin, we introduce a general WSLNR in pursuit of incorporating the notion of WGI as
\begin{equation}\label{eq:Xi_def}
    \chi_{i,t} =
    \begin{cases}
    \frac{\beta_{ii,t}\left|\mathbf{h}_{ii,t}\mathbf{w}_{i,t}\right|^2}{\sum_{j\in\mathcal{N}_{C}\setminus\{i\}}\beta_{ij,t}\left|\mathbf{h}_{ij,t}\mathbf{w}_{i,t}\right|^2+N_0} & \text{ if } \beta_{ii,t}\neq0\\
    \frac{1}{\sum_{j\in\mathcal{N}_{C}\setminus\{i\}}\beta_{ij,t}\left|\mathbf{h}_{ij,t}\mathbf{w}_{i,t}\right|^2+N_0}& \text{ if } \beta_{ii,t}=0,
    \end{cases}
\end{equation}
which is proposed in \cite{Y_Kim20_TVT}.
One of the key ideas of this letter is to limit the weight coefficients $\beta_{ij,t}$, $i,j\in\mathcal{N}_C$, to 0 or 1\footnote{In \cite{Y_Kim20_TVT}, it is shown that 0 and 1 are sufficient candidates for the weight coefficients in \eqref{eq:Xi_def} to maximize the sum-rate.} to reduce the required information exchange.
If $\beta_{ii,t}=1$, the beamforming vector of BS $i$ is designed as 
\begin{align}  \label{eq:maxWSLNR_w0}
    \mathbf{w}_{i,t} &= \arg \max_{\left\|\mathbf{w}\right\|^2=1} \frac{\left|\mathbf{h}_{ii,t}^{H}\mathbf{w}\right|^2}{\sum_{j\in\mathcal{N}_C\setminus\{i\}}\beta_{ij}\left|\mathbf{h}_{ij,t}^{H}\mathbf{w}\right|^2+N_0}\\
    &=  \arg \max_{\left\|\mathbf{w}\right\|^2=1} \frac{\mathbf{w}^{H}\mathbf{A}_{i,t}\mathbf{w}}{\mathbf{w}^{H}\mathbf{B}_{i,t}\mathbf{w}},
\end{align}
where $\mathbf{A}_{i,t}=\mathbf{h}_{ii,t}\mathbf{h}_{ii,t}^{H}$ and $\mathbf{B}_{i,t}={\displaystyle \sum_{j\in\mathcal{N}_C\setminus\{i\}}}\beta_{ij,t}\mathbf{h}_{ij,t}\mathbf{h}_{ij,t}^{H}+N_{0}\mathbf{I}$.
Then, the solution of \eqref{eq:maxWSLNR_w0} is given by the eigenvector of $\mathbf{B}_{i,t}^{-1}\mathbf{A}_{i,t}$ associated with the maximum eigenvalue.

If $\beta_{ii,t}=0$ and $\boldsymbol{\beta}_{i,t}\neq\mathbf{0}$, the beamforming vector of BS $i$ is designed as
\begin{align}
    \mathbf{w}_{i,t} & = \arg \max_{\left\|\mathbf{w}\right\|^2=1} \frac{1}{\sum_{j\in\mathcal{N}_C\setminus\{i\}}\beta_{ij,t}\left|\mathbf{h}_{ij,t}^{H}\mathbf{w}\right|^2+N_0}   \label{eq:w_i1} \\ 
    & = \arg \min_{\left\|\mathbf{w}\right\|^2=1} \left\Vert \mathbf{G}_{i,t}\mathbf{w}\right\Vert ^{2},
\end{align}
where $\mathbf{G}_{i,t} \triangleq \left[\sqrt{\beta_{i1,t}}\mathbf{h}_{i1,t}, \ldots, \sqrt{\beta_{iN_{C},t}}\mathbf{h}_{iN_{C},t}\right]^{H}$. Then, the solution for the problem \eqref{eq:w_i1} is obtained by choosing the right singular vector of $\mathbf{G}_{i,t}$ associated with the smallest singular value.
If $\boldsymbol{\beta}_{i,t}=\mathbf{0}$, the beamforming vector of BS $i$ is designed as
\begin{equation} \label{eq:w0}
    \mathbf{w}_{i,t}=\mathbf{0}.
\end{equation}
Then, under the limit of $\beta_{ij,t}\in\{0,1\}$, $i,j\in\mathcal{N}_C$, the number of possible cases of $\boldsymbol{\beta}_{i,t}$ designs for BS $i$ is $2^{N_C}$.
Now, the goal is to obtain $\boldsymbol{\beta}_{i,t}$, $i\in\mathcal{N}_C$, which maximizes the sum-rate with local CSI and a DQN in a distributed way.

\begin{center}
\begin{figure*}[ht]
\centering{}\includegraphics[width=\if 1\mycmd 1 \else 0.9 \fi \textwidth]{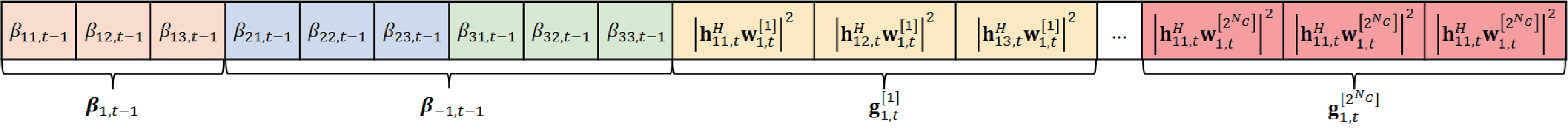}\caption{Example of the state of the proposed DQN algorithm of BS 1 over the time slot $t$ for $N_C=3$}\label{fig:state}
\end{figure*}
\par\end{center}

\section{Proposed DQN based beamforming}\label{sec:proposed_dqn}

In the proposed scheme, a DQN is utilized to choose the proper weight coefficients $\boldsymbol{\beta}_{i,t}$, $i\in\mathcal{N}_C$, for \eqref{eq:Xi_def} to maximize the sum-rate.
Each BS becomes the agent of the proposed DQN in turn and learns the DQN offline, which guarantees the reduced convergence time by sharing the policy information between all the BSs.
The state of the proposed DQN is composed of the weight coefficients of all the BSs and the scalar information, which can be calculated with only local CSI by the corresponding agent.
The action of each agent is to update the selection on the weight coefficients, and the reward is the sum-rate gain due to the update of action.
The proposed DQN is composed of 7 fully connected layers, and ReLu is used for the activation function.
In the following subsections, the elements of the proposed DQN algorithm is detailed.

\subsection{Definition of State, Action, and Reward}

\noindent\textbf{State:}
In the proposed network, the state of BS $i$ over the time slot $t$ is defined as
\begin{equation}                                          
    \mathbf{s}_{i,t}=\left[\boldsymbol{\beta}_{i,t-1},\boldsymbol{\beta}_{-i,t-1},\mathbf{g}_{i,t}^{[1]},\ldots,\mathbf{g}_{i,t}^{[2^{N_C}]}\right],
\end{equation}
where $\boldsymbol{\beta}_{-i,t}$ denotes the vector of $\boldsymbol{\beta}_{j,t}$, $j\neq i$, and $\mathbf{g}_{i,t}^{[c]}=\left[\left|\mathbf{h}_{i1,t}^H\mathbf{w}_{i,t}^{[c]}\right|^2,\ldots,\left|\mathbf{h}_{iN_C,t}^H\mathbf{w}_{i,t}^{[c]}\right|^2\right]$.
Fig. \ref{fig:state} shows the example of the state of BS 1 over the time slot $t$ for $N_C=3$. 
Thus, it can be seen that the state of the BS $i$ over the time slot $t$ consists of the weight coefficients of all the BSs from the very previous time slot $t-1$ and all the outgoing channel gains of BS $i$ according to all the possible actions for the present time slot $t$.
Note that the state of BS $i$, $\mathbf{s}_{i,t}$, is composed of the vectors which can be calculated with only local CSIT by BS $i$ except $\boldsymbol{\beta}_{-i,t}$ and the size of the state is $N_C (N_C+2^{N_C})$.

\vspace{0.3cm}
\noindent\textbf{Action:}
For the given state $\mathbf{s}_{i,t}$, the action of BS $i$ over the time slot $t$ and the action space are defined as $a_{i,t}$ and $\mathcal{A}$, respectively. In the proposed scheme, the action of BS $i$ is to update $\boldsymbol{\beta}_{i,t}$ to maximize the sum-rate. Thus, the action space $\mathcal{A}$ is represented as 
\begin{equation}
    \mathcal{A}=\{1,\ldots,2^{N_C}\}.
\end{equation}
In the exploration perspective, it is extremely important to reduce the action space size in RL applications for training time reduction as in \cite{G_Dulac16_arXiv}.
The training time of the proposed DQN is reduced by solving the classification problem with the finite action space size instead of the beamforming vector design problem which has the infinite action space size.

\vspace{0.3cm}
\noindent\textbf{Reward:}
The reward of the proposed DQN for the action $a_{i,t}$ given the state $\mathbf{s}_{i,t}$ is defined as
\if 1\mycmd
\begin{equation} \label{eq:reward}
    r_t = R_t \left( \boldsymbol{\beta}_{1,t-1}, \ldots, \boldsymbol{\beta}_{i,t}, \ldots, \boldsymbol{\beta}_{N_C,t-1} \right) -R_{t-1}\left( \boldsymbol{\beta}_{1,t-1}, \ldots, \boldsymbol{\beta}_{i,t-1}, \ldots, \boldsymbol{\beta}_{N_C,t} \right),
\end{equation}
\else
\begin{align} \label{eq:reward}
    r_t = & R_t \left( \boldsymbol{\beta}_{1,t-1}, \ldots, \boldsymbol{\beta}_{i,t}, \ldots, \boldsymbol{\beta}_{N_C,t-1} \right) \nonumber
    \\ & -R_{t-1}\left( \boldsymbol{\beta}_{1,t-1}, \ldots, \boldsymbol{\beta}_{i,t-1}, \ldots, \boldsymbol{\beta}_{N_C,t} \right),
\end{align}
\fi
which is the increase in the sum-rate after the update from $\boldsymbol{\beta}_{i,t-1}$ to $\boldsymbol{\beta}_{i,t}$.
By choosing the increase in the sum-rate after the update from $\boldsymbol{\beta}_{i,t-1}$ to $\boldsymbol{\beta}_{i,t}$ as the reward of the proposed DQN instead of the sum-rate $R_t \left( \boldsymbol{\beta}_{1,t-1}, \ldots, \boldsymbol{\beta}_{i,t}, \ldots, \boldsymbol{\beta}_{N_C,t-1} \right)$ itself, it is expected that the probability of the decrease in sum-rate compared to the previous time slot would be largely diminished.

\subsection{Proposed DQN Algorithm}

We propose the following DQN algorithms for the distributed beamforming vector design. Algorithm \ref{Algo:algorithm_dqn} shows the procedure of the proposed DQN. We have the following observations on Algorithm \ref{Algo:algorithm_dqn}.
\begin{itemize}
\item Only one BS updates its weight coefficients at a time slot, based on the scalar information which can be calculated with only local CSI and the weight coefficients of the other BSs at the previous time slot. As a result, only the weight coefficients of the BS which is the previous agent are needed to be shared.
\item For the proposed DQN, the reward is not needed to be calculated since the DQN doesn't have to be updated for the implementation procedure. Thus, the information exchange for the sum-rate calculation is not required for the implementation of the proposed DQN.
\end{itemize}

\begin{algorithm}[ht]
\caption{The algorithm process for the proposed DQN}\label{Algo:algorithm_dqn}
\begin{algorithmic}[1]
\State Initialization ($\boldsymbol{\beta}_{i,0}=\boldsymbol{1}$, for all $i\in\mathcal{N}_C$)
\For{$t=0$ to $T$} \Comment{$T$: The maximum time slot}
    \State $i\leftarrow(t\mod{N_C})$ \Comment{The agent index at $t$}
    \State BS $i$ becomes the agent for the time slot $t$
    \If{$t>0$}
        \State $j\leftarrow((t-1)\mod{N_C})$ \Comment{The agent index at $t-1$}
        \State BS $j$ shares $\boldsymbol{\beta}_{j,t-1}$ with BS $i$ for the state $\mathbf{s}_{i,t}$
    \EndIf
    \If{training mode}
        \State BS $i$ trains the DQN
    \EndIf
    \State Update $\boldsymbol{\beta}_{i,t}$ according to the proposed DQN
    \State Calculate the reward $r_{t}$ as \eqref{eq:reward}
\EndFor
\end{algorithmic}
\end{algorithm}

Fig. \ref{fig:algorithm} shows the example of the detailed procedure of the proposed DQN for $N_C=3$, which is introduced in Algorithm \ref{Algo:algorithm_dqn}. At the time slot $t-1$, the agent is BS 3 and the weight coefficients of the BS 3 is updated with the action $a_{3,t-1}$. Then, the sum-rate is updated from $R_{t-2}(\boldsymbol{\beta}_{1,t-2},\boldsymbol{\beta}_{2,t-2},\boldsymbol{\beta}_{3,t-2})$ to $R_{t-1}(\boldsymbol{\beta}_{1,t-2},\boldsymbol{\beta}_{2,t-2},\boldsymbol{\beta}_{3,t-1})$ with the updated weight coefficients $\boldsymbol{\beta}_{3,t-1}$. At the time slot $t$ and $t+1$, BS 1 and BS 2 become the agent of the DQN, respectively, and they update their weight coefficients.

\begin{center}
\begin{figure}[ht]
\centering{}\includegraphics[width=\if 1\mycmd 0.7 \else 0.48 \fi \textwidth]{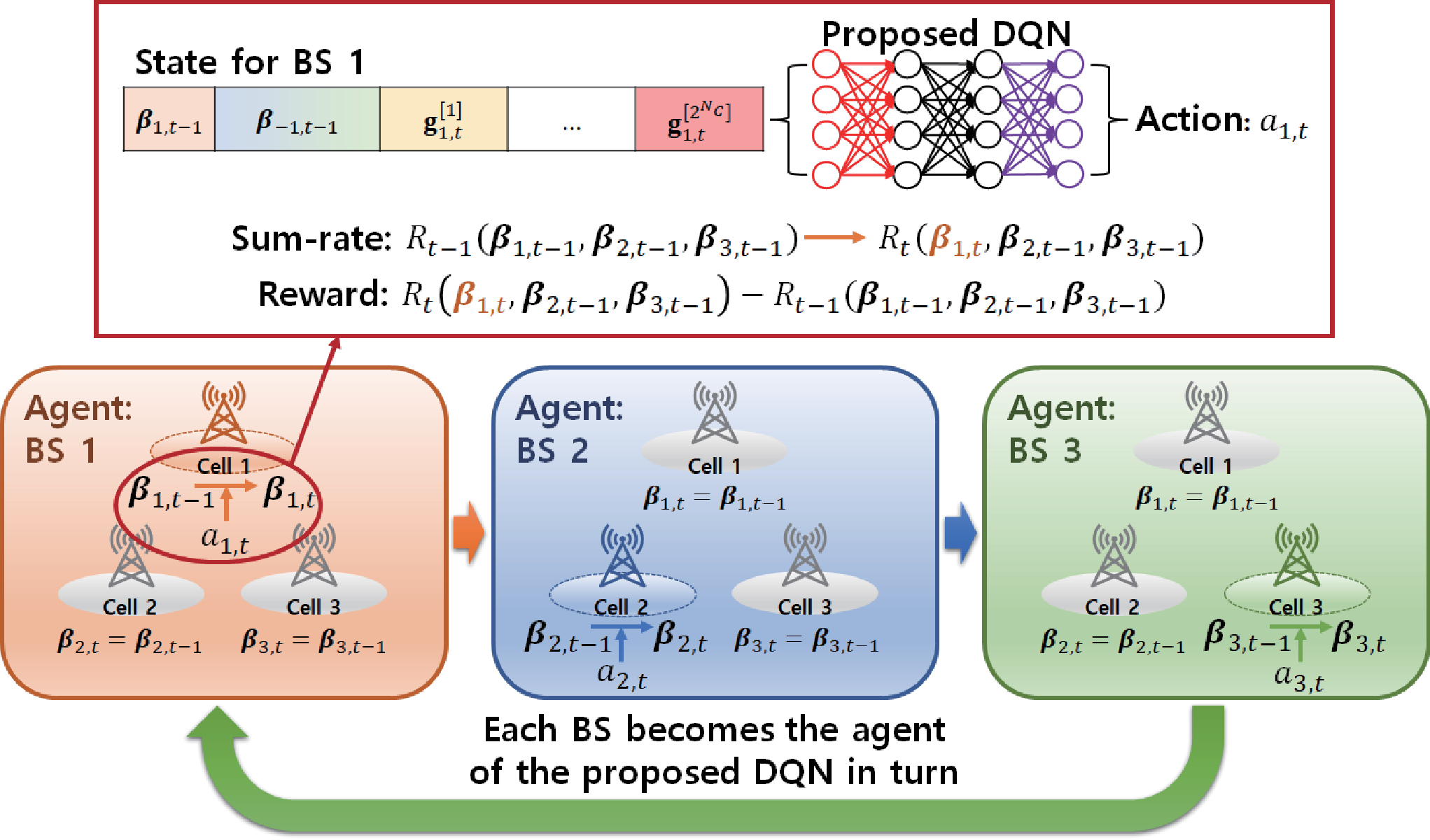}\caption{Example of the proposed DQN algorithm for $N_C=3$}\label{fig:algorithm}
\end{figure}
\par\end{center}

\subsection{Information Exchange}\label{sec:informaiton_exchange}

In this subsection, the amount of information exchange required in the proposed scheme is compared to that of the existing scheme.
The scheme, which maximizes the sum-rate by choosing the proper set of interference-free users \cite{Y_Kim20_TVT}, is considered, where only limited scalar information exchange is required. The scheme is referred to as `IFU-selection' henceforth. 

For the state, which is the input of the proposed DQN, $\boldsymbol{\beta}_{-i,t}$ term needs to be shared via a direct link with limited capacity to BS $i$ for each time slot.
Let assume that BS $j$ is the agent of the previous time slot and $\boldsymbol{\beta}_{j,t-1}$ was updated to $\boldsymbol{\beta}_{j,t}$.
Then, for the present time slot, the only information about $\boldsymbol{\beta}_{j,t}$ needs to be shared with the present agent.
Thus, only $N_C$ bits of information exchange are required per each time slot for the proposed DQN algorithm.
Table \ref{table:information_exchange} summarizes the amount of required information exchange in bits for the proposed scheme and `IFU-selection' \cite{Y_Kim20_TVT}.

\begin{table*}[tbh]
\caption{Amount of required information exchange per each time slot (in bits)}\label{table:information_exchange}
\begin{center}
\begin{tabular}{cccccc}
    \cmidrule[\heavyrulewidth]{1-6}
     & \textbf{General formulation} & \multicolumn{4}{c}{\textbf{Amount of information exchange (bits)}} \\ \cmidrule(r){2-2} \cmidrule(r){3-6}
     &  \makecell[c]{$N_T$: \# of Tx antennas, $N_C$: \# of cells\\ $N_f$: The amount of required \\ information exchange by each BS in \cite{Y_Kim20_TVT}} & \makecell{$N_T=3$, $N_C=6$,\\ $N_f=3$} &  \makecell{$N_T=3$, $N_C=7$,\\ $N_f=5$} &  \makecell{$N_T=4$, $N_C=6$,\\ $N_f=42$} & \makecell{$N_T=4$, $N_C=7$,\\ $N_f=42$} \\ \midrule
     \textbf{Proposed scheme} & $N_C$ & \textbf{6} & \textbf{7}  & \textbf{6} & \textbf{7} \\[7pt] 
     \textbf{IFU-selection \cite{Y_Kim20_TVT}} & $(N_C-1)\left(N_f+\left\lceil\log\sum_{\alpha = 1}^{N_T}\binom{N_C}{\alpha} \right\rceil\right)$ & 110 & 168 & 160 & 294\\[2pt]
     \bottomrule
\end{tabular}
\end{center}
\end{table*}



\section{Numerical Simulations}
For the numerical simulation, the proposed DQN is trained for 20,000 episodes by circularly using 5000 training data and is evaluated for another 20,000 episodes with 20,000 test data.
Table \ref{table:parameter} summarizes the network parameters which are used for the simulations.
The proposed DQN is implemented by Python 3.8 with Pytorch 1.11.0 library and Numpy library.
For all the numerical simulations in this section, small cell networks with time-varying channels are considered \cite{TR25.996,TR36.814,TR36.931}.

\begin{figure}[tbh]
  \centering
  \subfloat[$N_T=3$, $N_C=6$ \label{fig:step_Nt3Nc6}]{\includegraphics[width=\if 1\mycmd 0.5 \else 0.25 \fi \textwidth]{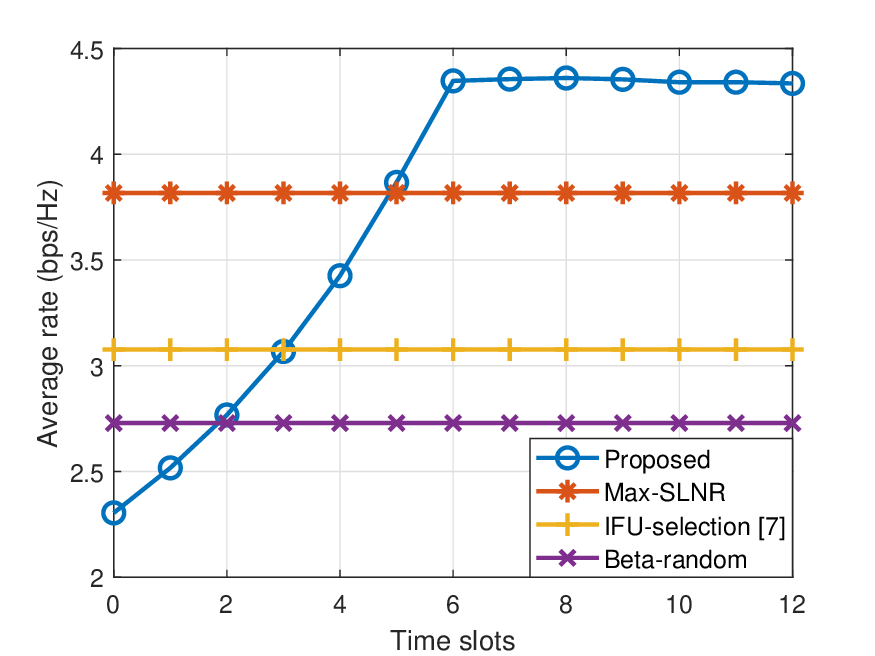}}
  \subfloat[$N_T=3$, $N_C=7$ \label{fig:step_Nt3Nc7}]{\includegraphics[width=\if 1\mycmd 0.5 \else 0.25 \fi\textwidth]{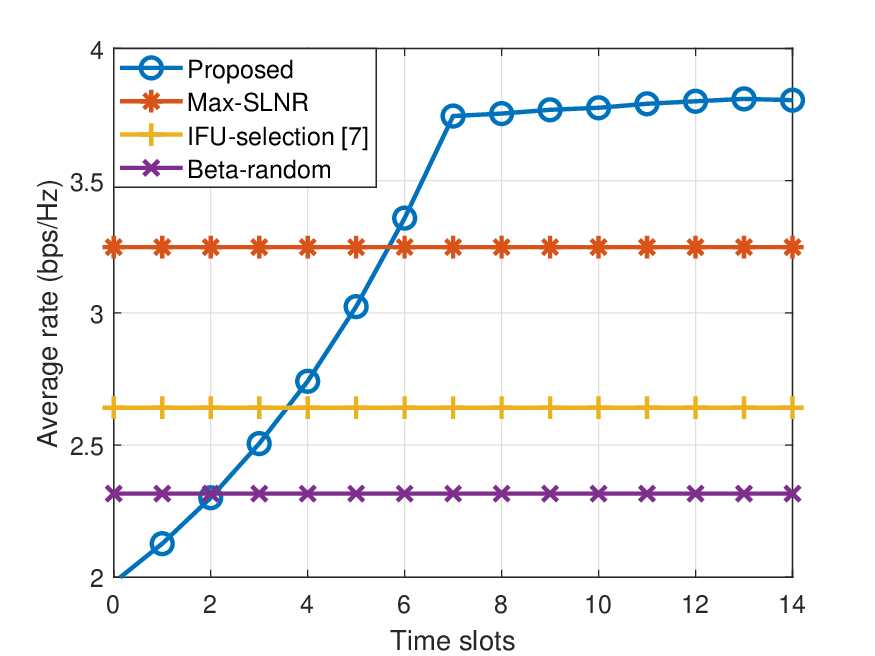}}\\
  \subfloat[$N_T=4$, $N_C=6$ \label{fig:step_Nt4Nc6}]{\includegraphics[width=\if 1\mycmd 0.5 \else 0.25 \fi\textwidth]{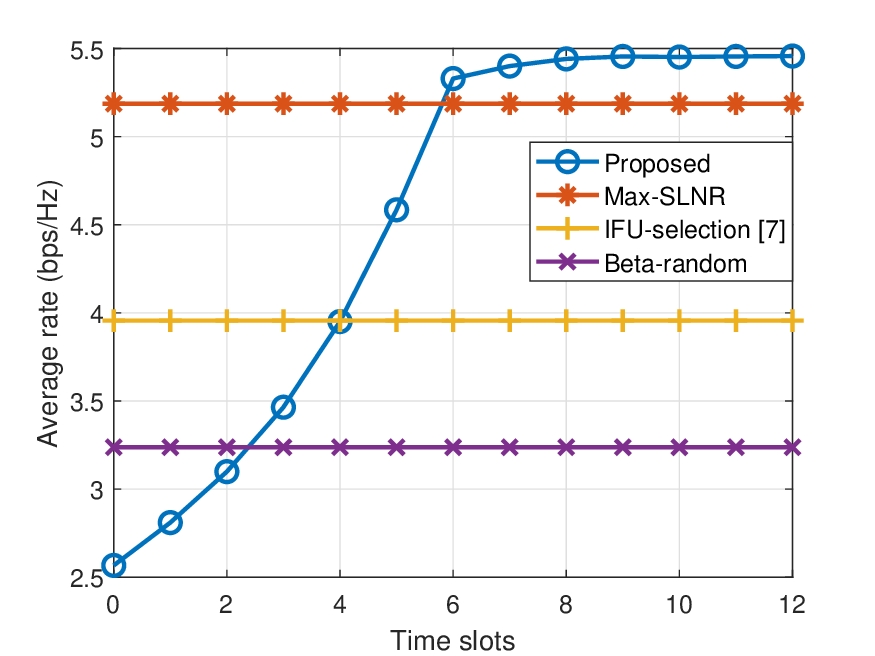}}
  \subfloat[$N_T=4$, $N_C=7$ \label{fig:step_Nt4Nc7}]{\includegraphics[width=\if 1\mycmd 0.5 \else 0.25 \fi\textwidth]{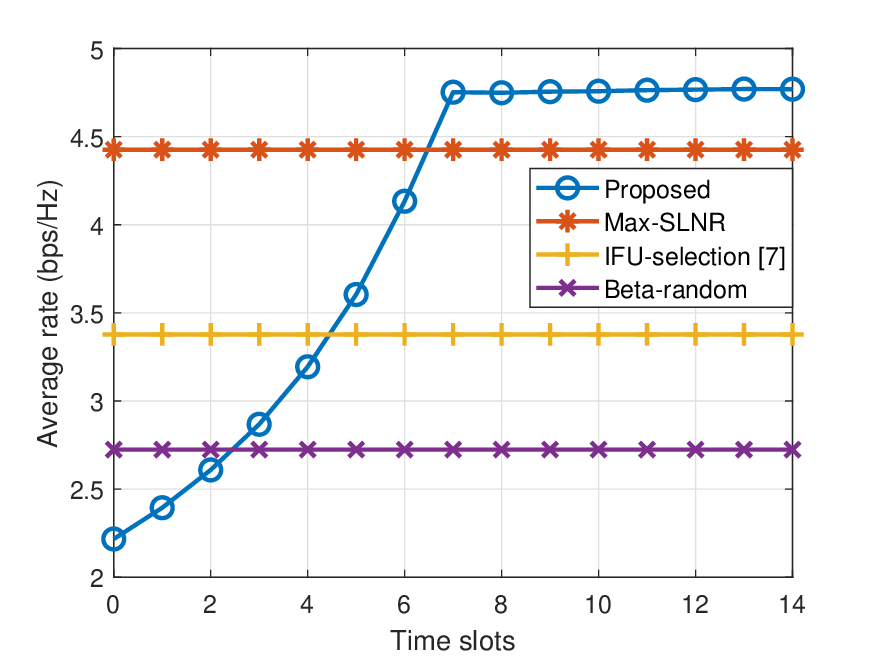}}
  \caption{Per-cell average rate versus time slot \label{fig:step}.}
\end{figure}

\begin{figure}[tbh]
  \centering
  \subfloat[$N_T=3$, $N_C=6$ \label{fig:sumrate_Nt3Nc6}]{\includegraphics[width=\if 1\mycmd 0.5 \else 0.25 \fi\textwidth]{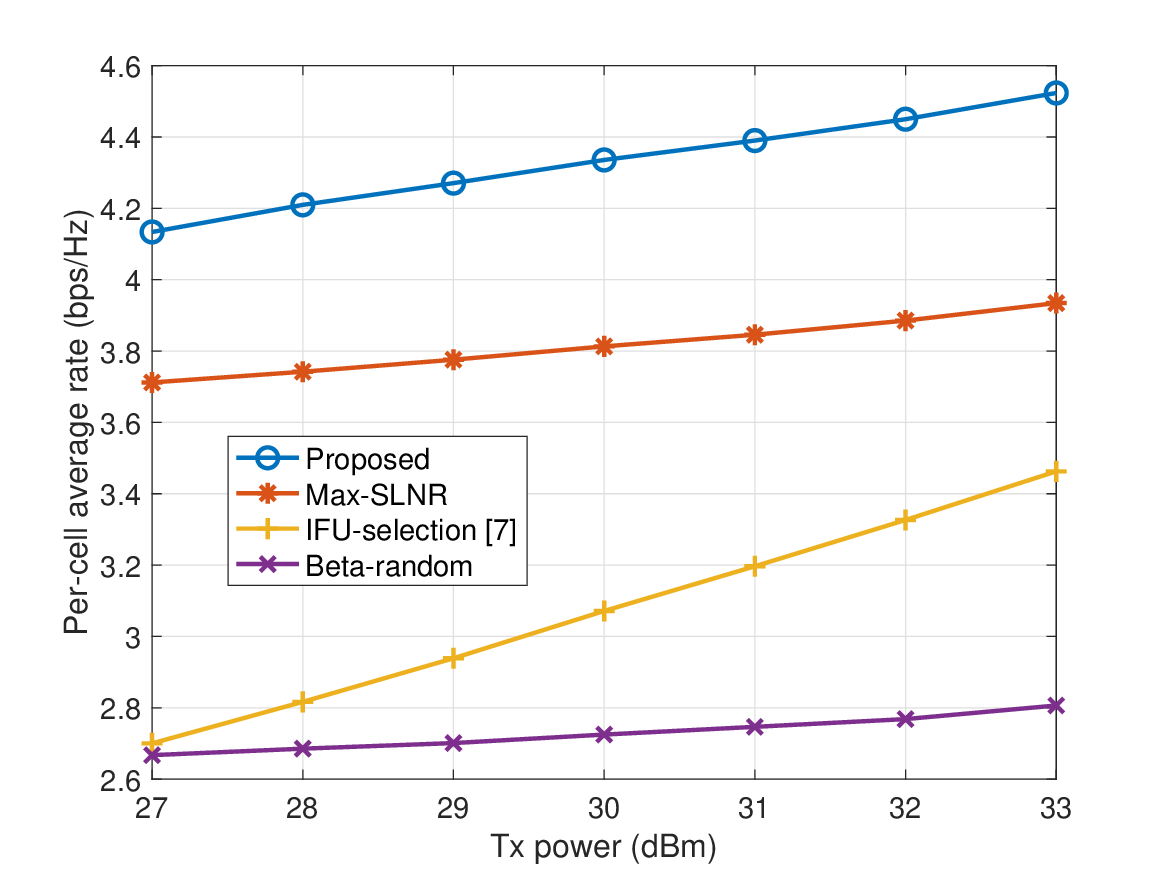}} 
  \subfloat[$N_T=3$, $N_C=7$ \label{fig:sumrate_Nt3Nc7}]{\includegraphics[width=\if 1\mycmd 0.5 \else 0.25 \fi\textwidth]{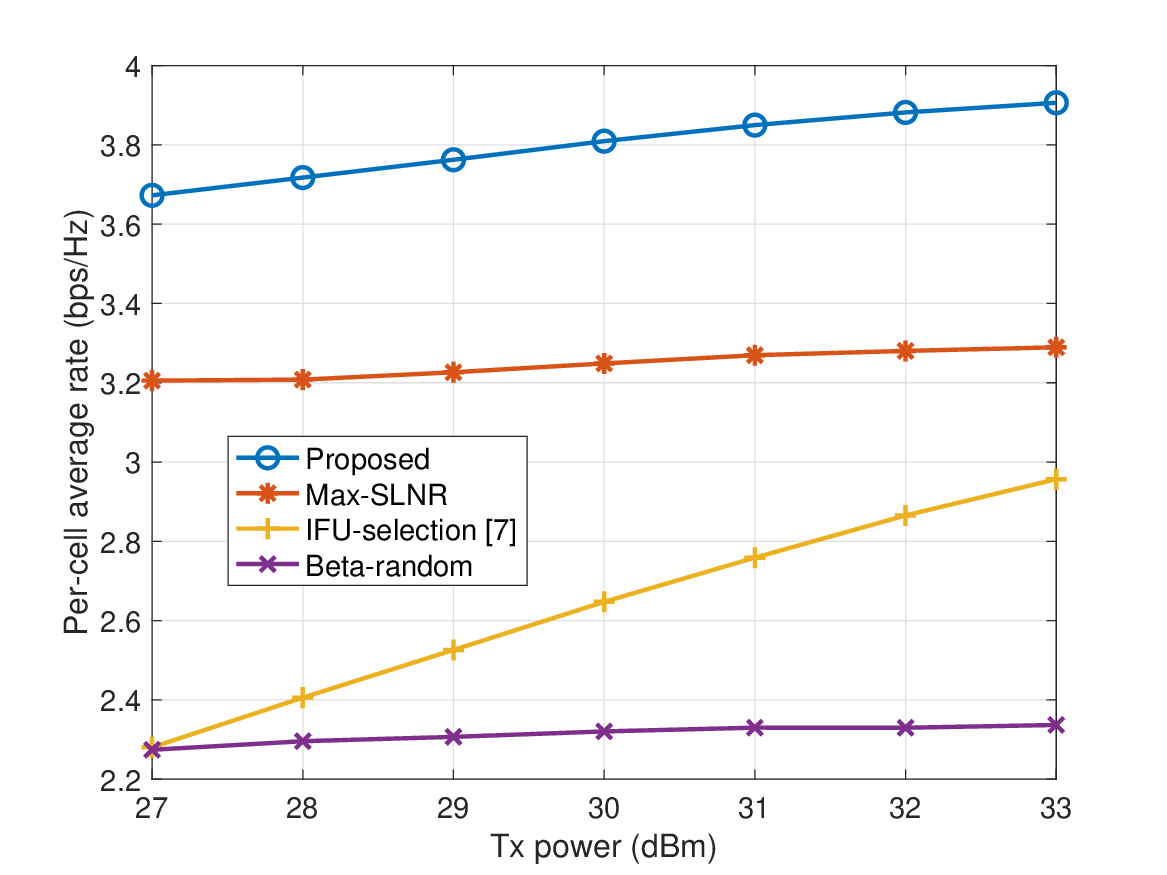}} \\
  \subfloat[$N_T=4$, $N_C=6$ \label{fig:sumrate_Nt4Nc6}]{\includegraphics[width=\if 1\mycmd 0.5 \else 0.25 \fi\textwidth]{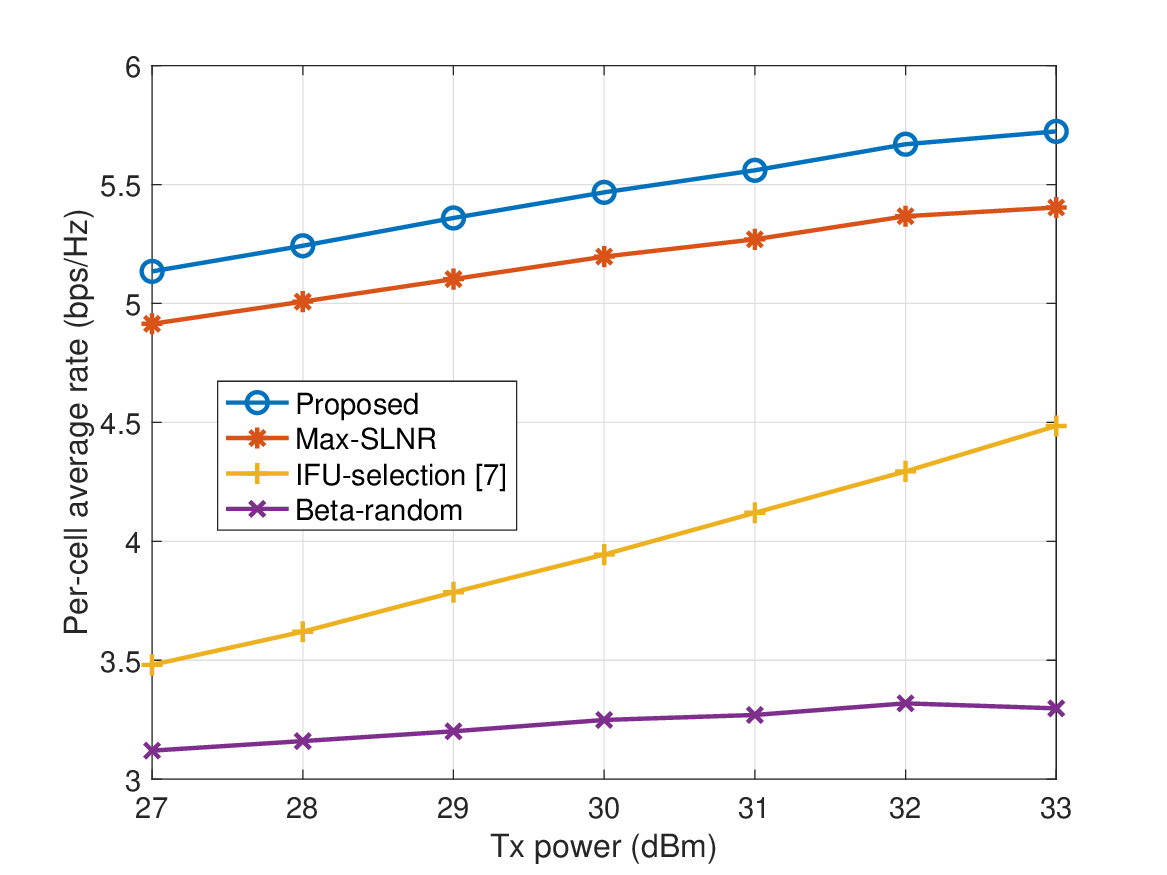}}
  \subfloat[$N_T=4$, $N_C=7$ \label{fig:sumrate_Nt4Nc7}]{\includegraphics[width=\if 1\mycmd 0.5 \else 0.25 \fi\textwidth]{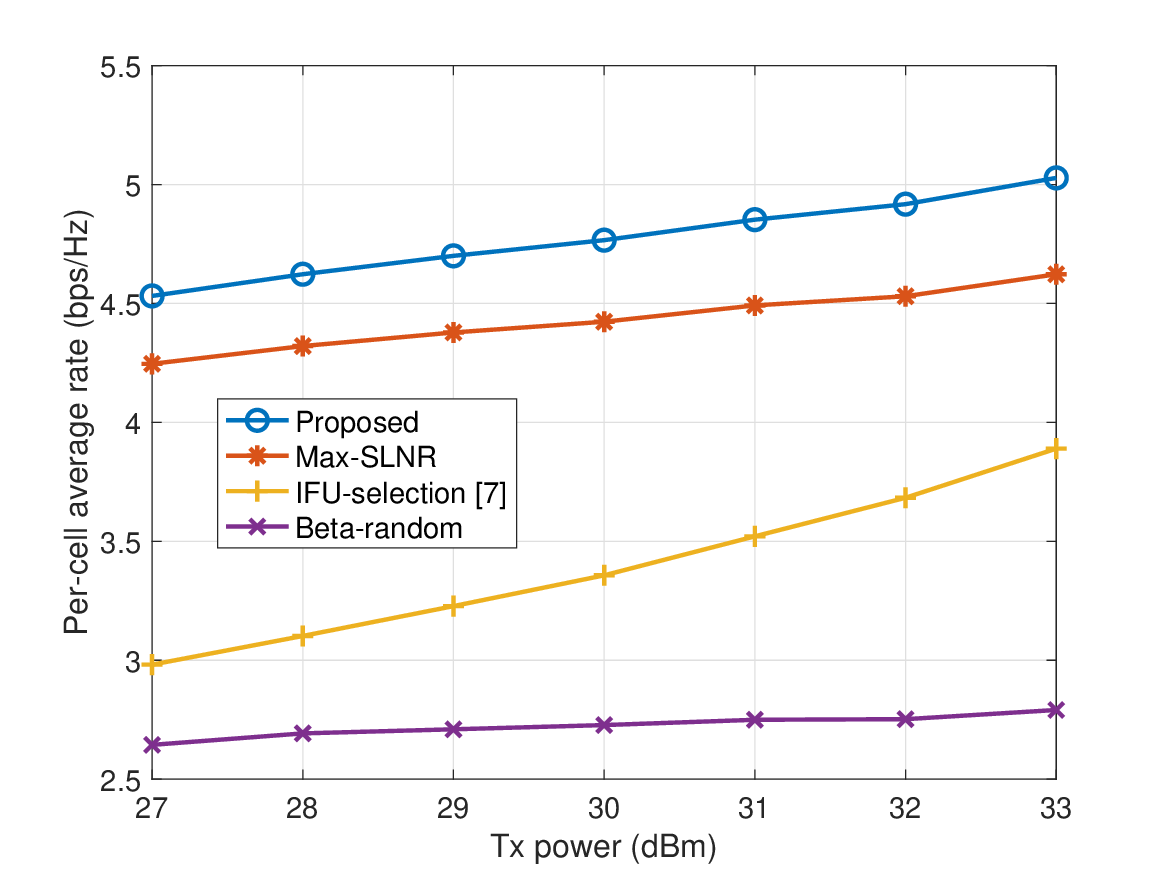}}
  \caption{Per-cell average rate versus transmit power of BSs\label{fig:sumrate}.}
\end{figure}

\begin{figure}[tbh]
  \centering
  \subfloat[$N_T=3$, $N_C=6$ \label{fig:info_Nt3Nc6}]{\includegraphics[width=\if 1\mycmd 0.5 \else 0.25 \fi\textwidth]{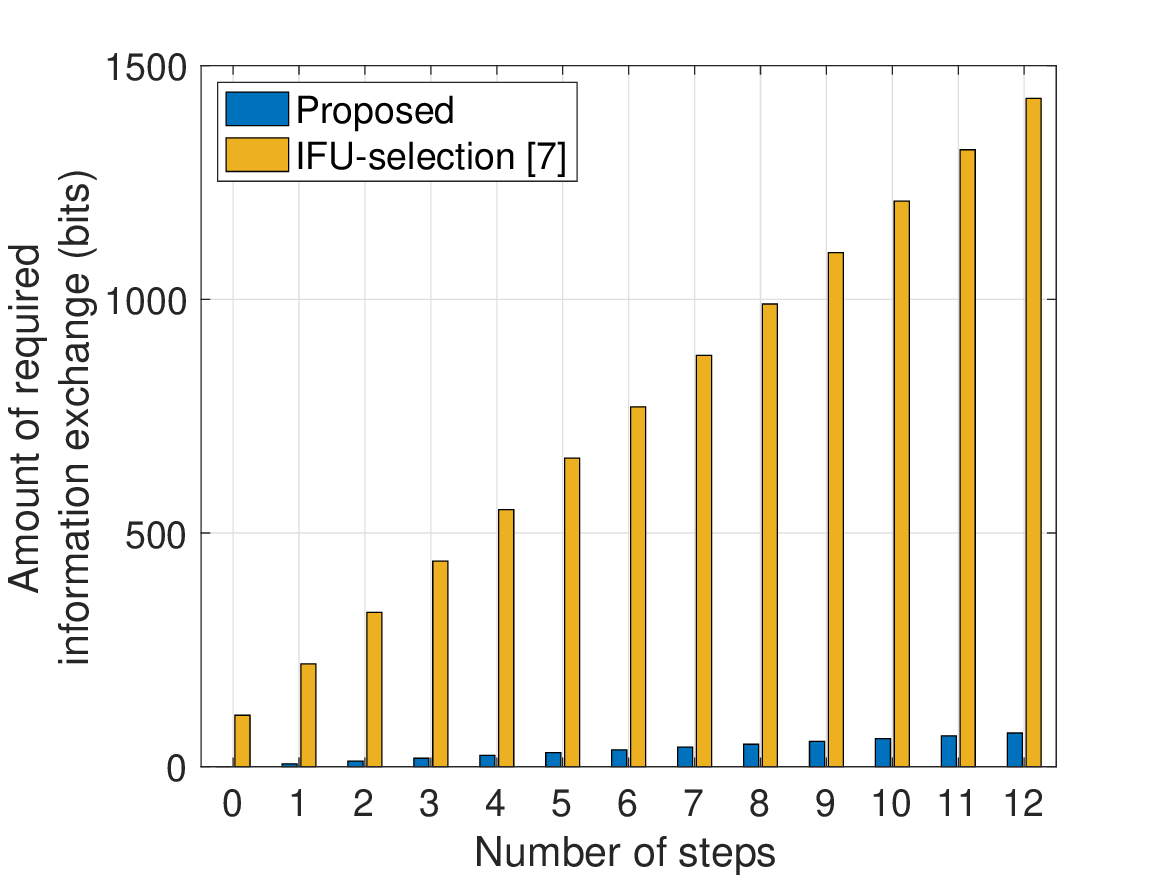}}
  \subfloat[$N_T=4$, $N_C=7$ \label{fig:info_Nt4Nc7}]{\includegraphics[width=\if 1\mycmd 0.5 \else 0.25 \fi\textwidth]{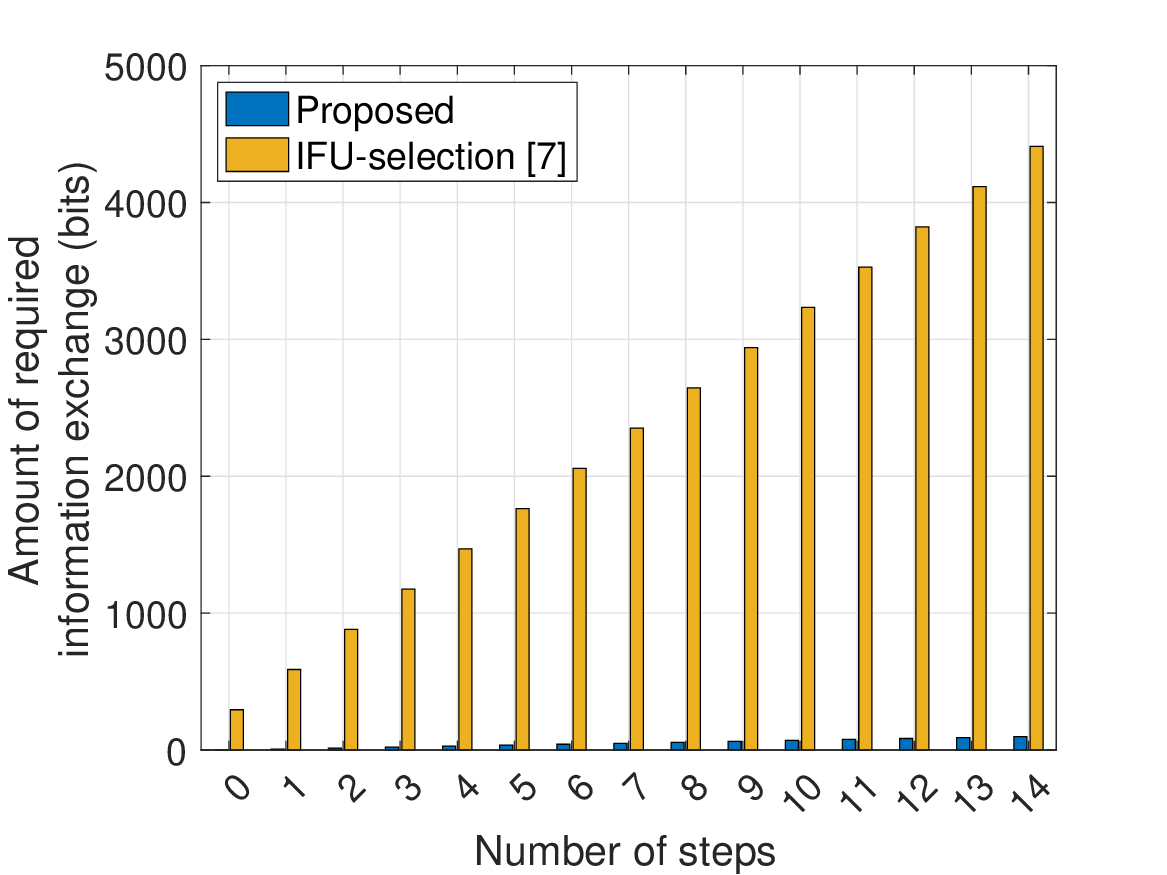}}
  \caption{The required information exchange in bits\label{fig:info}.}
\end{figure}

In Figs. \ref{fig:step} and \ref{fig:sumrate}, the per-cell average rate of the proposed scheme is compared with some existing schemes for four scenarios: ($N_T=3$, $N_C=6$), ($N_T=3$, $N_C=7$), ($N_T=4$, $N_C=6$) and ($N_T=4$, $N_C=7$).
Here, `IFU-selection' which is discussed in Section \ref{sec:informaiton_exchange}, `Max-SLNR,' and `Beta-random' are considered.
In `Max-SLNR,' all the BSs construct their beamforming vectors maximizing SLNR.
In `Beta-random,' the beamforming vectors are designed to maximize \eqref{eq:Xi_def}, where all the weight coefficients in \eqref{eq:Xi_def} are generated randomly.
For Figs. \ref{fig:step} and \ref{fig:sumrate}, the x-axes are the time slot and the transmit power of BS, respectively.
In Figs. \ref{fig:info_Nt3Nc6} and \ref{fig:info_Nt4Nc7}, the amounts of required information exchange for the proposed scheme and `IFU-selection' are compared.

As shown in Fig. \ref{fig:step}, the proposed scheme shows the highest per-cell average rate among the considered scheme after $N_C$ time slots, showing about 5$\sim$17\% improved per-cell average rates over `Max-SLNR'. That is, with only one beamforming vector update for all the BSs, the proposed scheme can achieve notable per-cell average rate gain. Compared to `IFU-selection,' it can be seen that the gap between the amounts of required information exchange of the proposed scheme and `IFU-selection' increases over time as shown in Fig \ref{fig:info}, while the performance gap between the per-cell average rates of the proposed scheme and `Max-SLNR' is maintained after $N_C$ time slots as shown in Fig. \ref{fig:step}.

As shown in Fig. \ref{fig:sumrate}, the proposed scheme achieves the highest per-cell average rate among the considered schemes.
In particular, the proposed scheme shows much higher per-cell average rates compared to `IFU-selection' with just $N_C$ bits of information exchange per each time slot as shown in Table \ref{table:information_exchange}.
Also, the gap between the per-cell average rate of the proposed scheme and that of `Max-SLNR' increases as the transmit power increases, since the importance of intercell interference management with the proper weight coefficients in \eqref{eq:Xi_def} increases as the overall SNR increases with transmit power.
The gap between the per-cell average rate of the proposed scheme and that of `Beta-random' shows that the proposed scheme can achieve notable per-cell average rate gain by choosing the proper weight coefficients with the efficiently trained proposed DQN.

\begin{table}[tbh]
\caption{Network parameters used in the simulation \cite{TR25.996,TR36.814,TR36.931}}\label{table:parameter}
\begin{center}
\begin{tabular}{ll}
\toprule
\textbf{Parameter} & \textbf{Value}                \\ \cmidrule(r){1-1} \cmidrule(r){2-2}
Velocity of users (km/h) & 5                       \\ 
Cell radius (m)       & 70                         \\ 
Transmission power (dBm) & 27--33                    \\
Noise power (dBm/Hz)       & -174                  \\ 
Pathloss model (dB)    & 34.53 + 38$\log_{10}(d)$  \\ 
Learning rate      & 0.003                         \\ 
Gamma              & 0.99                           \\ \bottomrule
\end{tabular}
\end{center}
\end{table}

\section{Conclusion}
We have proposed the beamforming vector design for multicell MISO time-varying channels based on DQN.
We have designed an efficient algorithm to utilize the DQN to find the weight coefficients of WSLNR to maximize the sum-rate in a distributed way with a few scalar information exchange.
The proposed scheme can achieve notable sum-rate gain though it requires only $N_C$ bits of information exchange per each time slot.

Since only one user in each cell can be served by the corresponding BS in the proposed scheme, the other users inevitably have zero achievable rate.
In future work, the sum-rate maximizing beamforming optimization problem will be considered for multiuser per cell networks considering fairness between the users.

\bibliographystyle{IEEEtran}
\bibliography{BNN.bib}

\end{document}